\documentclass[a4paper,12pt]{article}
\usepackage{fullpage}
\usepackage{times}

\usepackage{authblk}

\usepackage{setspace}
\usepackage{caption}
\usepackage[utf8]{inputenc}
\usepackage{graphicx}
\graphicspath{ {./} }

\usepackage{newtxmath}
\usepackage{xcolor}
\usepackage{color}
\usepackage[normalem]{ulem}
\usepackage{lineno}

\usepackage[backend=biber,style=nature, doi=false,isbn=false,url=false,eprint=false,date=year]{biblatex}
\AtEveryBibitem{\clearfield{note}\clearfield{addendum}\clearlist{language}}
\AtEveryCitekey{\clearfield{note}\clearfield{addendum}\clearlist{language}}
\usepackage{titlesec}
\titleformat{\section}
  {\normalfont\bfseries}{\thesection}{1em}{}
\defbibheading{bibliography}[\refname]{%
  \section*{#1}%
  \markboth{#1}{#1}}

\usepackage[warnundef]{jabbrv}
\DefineJournalAbbreviation{Reports}{Rep}
\DefineJournalAbbreviation{physical}{Phys}
\DefineJournalAbbreviation{journal.}{J}

\begin{document}

\title{
Dynamic constraints predict the relaxation of granular materials
}

\author[1,2\dag]{Qinghao Mao}

\author[3,4,5\ddag]{Yujie Wang}

\author[6,*]{Walter Kob}

\affil[1]{Department of Physics, University of Chicago, Chicago, 60637, IL, USA}
\affil[2]{James Franck Institute, University of Chicago, Chicago, 60637, IL, USA}

\affil[3]{State Key Laboratory of Geohazard Prevention and Geoenvironment Protection, Chengdu University of Technology, Chengdu 610059, China}
\affil[4]{Department of Physics, College of Mathematics and Physics, Chengdu University of Technology, Chengdu 610059, China}
\affil[5]{School of Physics and Astronomy, Shanghai Jiao Tong University, Shanghai 200240, China}
\affil[6]{Department of Physics, University of Montpellier and CNRS, F-34095 Montpellier, France 
\vspace*{80mm} 
\newline
\hspace*{-42mm} $\dag$~qinghaomao@uchicago.edu\newline
\hspace*{-45mm} $\ddag$ yujiewang@sjtu.edu.cn\newline
\hspace*{-95mm} $^*$ walter.kob@umontpellier.fr}

\maketitle

\vspace{15mm}

\newpage

{\bf Abstract}
Granular materials such as sand, powders, and food grains are ubiquitous in civil engineering, geoscience, agriculture, and medicine.
While the influence of friction between the grains on the static structure of these systems is well understood, its impact on the dynamics is an open problem. Here we use particle-based simulations of a granular pack under cyclic shear and discover that the relaxation time of the system is a non-monotonic function of friction. By introducing the concept of dynamic constraints, we reveal that this re-entrant dynamics is due to the competition between increasing frictional coupling and a concurrent change in the structure of the granular pack. Our theoretical approach, which unifies the dynamics of friction-less systems with frictional ones, is applicable to other systems that have a complex free energy landscape and a dynamics which involves time-dependent constraints, thus setting the stage for a description of the dynamic behavior of a large class of complex systems. \\[5mm]

Due to the importance of granular materials for our daily life, they have been in the focus of interest of a multitude of studies~\cite{
andreotti2013,franklin2016,midi2004EPJESM,michlmayr2012ESR,behringer2018RPP}.
Critical insight into their static properties has come from 
various approaches, including Maxwell's method which connects mechanical constraints with the rigidity of a structure~\cite{maxwell1864,phillips1985SSC,wyart2005ADP,zaccone2011PRB,shundyak2007PRE},  force chain networks~\cite{snoeijer2004PRL,kondic2012EPL,abedzadeh2019GM,bi2011NATURE}, as well as the Edwards ensemble~\cite{edwards1989PHYSICAA, baule2018RMP, yuan2021PRL,xing2024NP}, which allows a statistical mechanics description of the  structure. 
Despite these successes, our understanding of the microscopic dynamics of granular packs, which features limit cycles, caging, convection, and a strong dependence on the history and driving protocol~\cite{silbert2002PRE,tordesillas2012PRE,kou2017NATURE,zhao2022PRX, royer2015PNAS, degiuli2016PRE,otsuki2021EPJE,sakellariou2017,lemaitre2021PRL},
remains unsatisfactory \cite{forterre2008ARFM}. This lack of insight is in contrast to the case of thermal glass-formers, which have a structure similar to the one of granular systems but whose relaxation dynamics is described reasonably well by multiple 
approaches~\cite{binder2011,gotze2008,lubchenko2007ARPC,dyre2006RMP}. One reason for this striking difference is that particle friction renders granular systems dynamically constrained and dissipative~\cite{wolf1998PM,silbert2002PRE,peshkov2019PRE,benson2022PRL}, hence requiring non-equilibrium approaches~\cite{franklin2016}. 

Theoretical studies of model glass-formers have revealed that dynamical slowing down can be caused by the presence of constraints~\cite{fredrickson1984PRL,kob1993PRE,ritort2003AIP}. These findings and the fact that the rigidity of granular systems can be explained by constraint counting~\cite{liu2021PRL,babu2023SM,vandernaald2024NP} 
hint that this approach can be generalized to describe the dynamics of driven granular materials. Here we probe this dynamics and demonstrate that dynamic constraints indeed allow to rationalize it. We argue that this approach is quite general and hence can be applied to a multitude of disordered many-body systems~\cite{charbonneau2023,hannezo2022TCB,atia2018NP,bigras2008DO}.\\

\begin{figure}[t!]
\includegraphics[width=12cm]{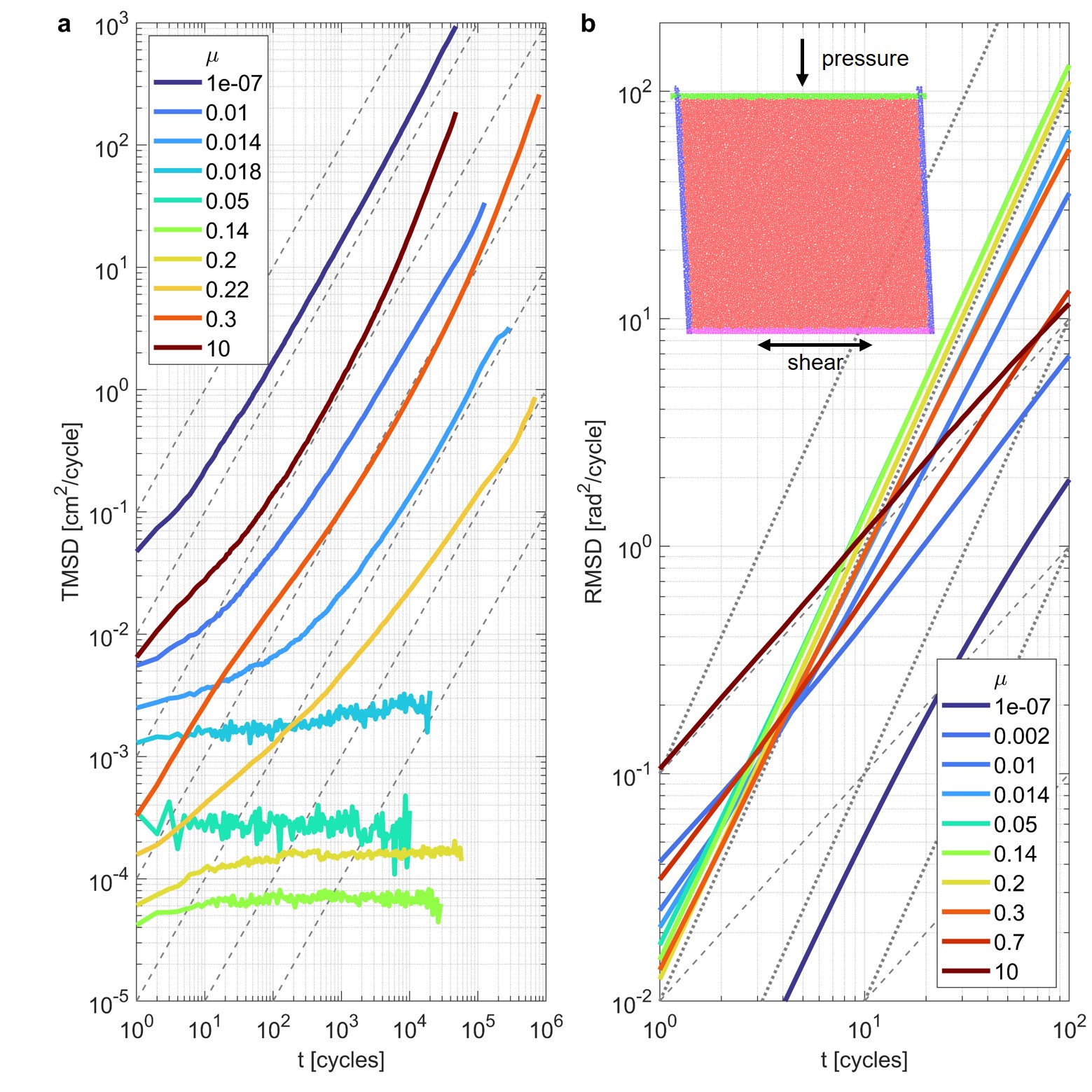}
\caption{
{\bf Translational and rotational dynamics:}
Time dependence of the translational mean squared displacement (TMSD), (a), and of the rotational one (RMSD), (b), for different friction constants $\mu$ (given in the legend). The dashed and dotted lines are power-laws with exponent 1.0 and 2.0, respectively.
The RMSD is obtained by time-integrating the angular displacements during the propagation of the system. Note that the value of these MSD's at $t=1$ is non-monotonic as a function of $\mu$, i.e., the dynamics is re-entrant. Depending on the value of $\mu$, the long time dynamics can be diffusive, super-diffusive, or caged.
The Inset in (b) shows the setup of the simulation. See Methods for details.
}
\label{fig_ED_TMSD_RMSD}
\end{figure}

{\bf System:} 
We investigate a 2:1 mixture of 2D frictional Hertzian particles having a diameter $d_s = \text{1~cm}$ (our unit of length) and $d_b=\text{1.4~cm}$, and a density $\rho = 1 \text{g} \cdot \text{cm}^{-2}$. 
The tangential component of the force, ${\bf F}_t$, is restricted by the Coulomb cone $|{\bf F}_t| \leq \mu |{\bf F}_n|$, where $\mu$ is the friction coefficient and ${\bf F}_n$ is the normal force at the contact. Details of the model and the simulations are given in the Methods. The total number of particles is $N=10^4$, and they undergo cyclic shear at constant pressure with a strain amplitude of 0.05 in a box of size 120~cm~$\times$~120~cm consisting of four rigid walls made of particles, see Inset of Fig.~\ref{fig_ED_TMSD_RMSD}b. In the following, time will be given in number of shear cycles.\\

{\bf Re-entrant dynamics:}
Figure~\ref{fig_ED_TMSD_RMSD} shows the time dependence of the translational and rotational mean squared displacements, TMSD and RMSD, respectively. The TMSD demonstrates that at small $\mu$ the dynamics has a fast, diffusion-like behavior while an increase of $\mu$ to 0.014 leads to a motion in which the particles are temporarily caged before they finally start to diffuse. For somewhat larger frictions the TMSD is small and basically constant, i.e.,~the particles are completely caged. This $\mu-$dependence is thus reminiscent of the slowing down of thermal glass-forming systems if the temperature is lowered~\cite{binder2011}, except that here it is the frictional coupling between the particles that is the control parameter. Surprisingly one finds that if friction is increased beyond $\mu \approx 0.2$, the dynamics speeds up again and at the highest $\mu$ the TMSD becomes hyperdiffusive. This re-entrant phenomenon, which is in qualitative agreement with the results of Ref.~\cite{royer2015PNAS} for a 3D system, hints that the relaxation dynamics is controlled by two competing processes that depend on $\mu$ and the nature of which will be elucidated below. 

The RMSD shows that at very small $\mu$ the rotational motion of the particles is slow and ballistic, i.e.,~increases quadratically with time, a result that is reasonable since the low friction and the slow shearing makes it hard for a particle to change its angular velocity. With growing $\mu$ the RMSD at short times increases quickly and its slope in this double logarithmic representation decreases to 1, indicating a rotational diffusive dynamics. For $\mu$ larger than $\approx 2\cdot10^{-3}$ this dynamics slows down again while its time-dependence turns back to ballistic. Finally, when $\mu$ is larger than $\approx 0.2$, the rotational movement speeds up again and becomes diffusive. Hence we conclude that also the rotational degrees of freedom show a re-entrant behavior as a function of friction.

\begin{figure}[ht]
\includegraphics[width=14cm]{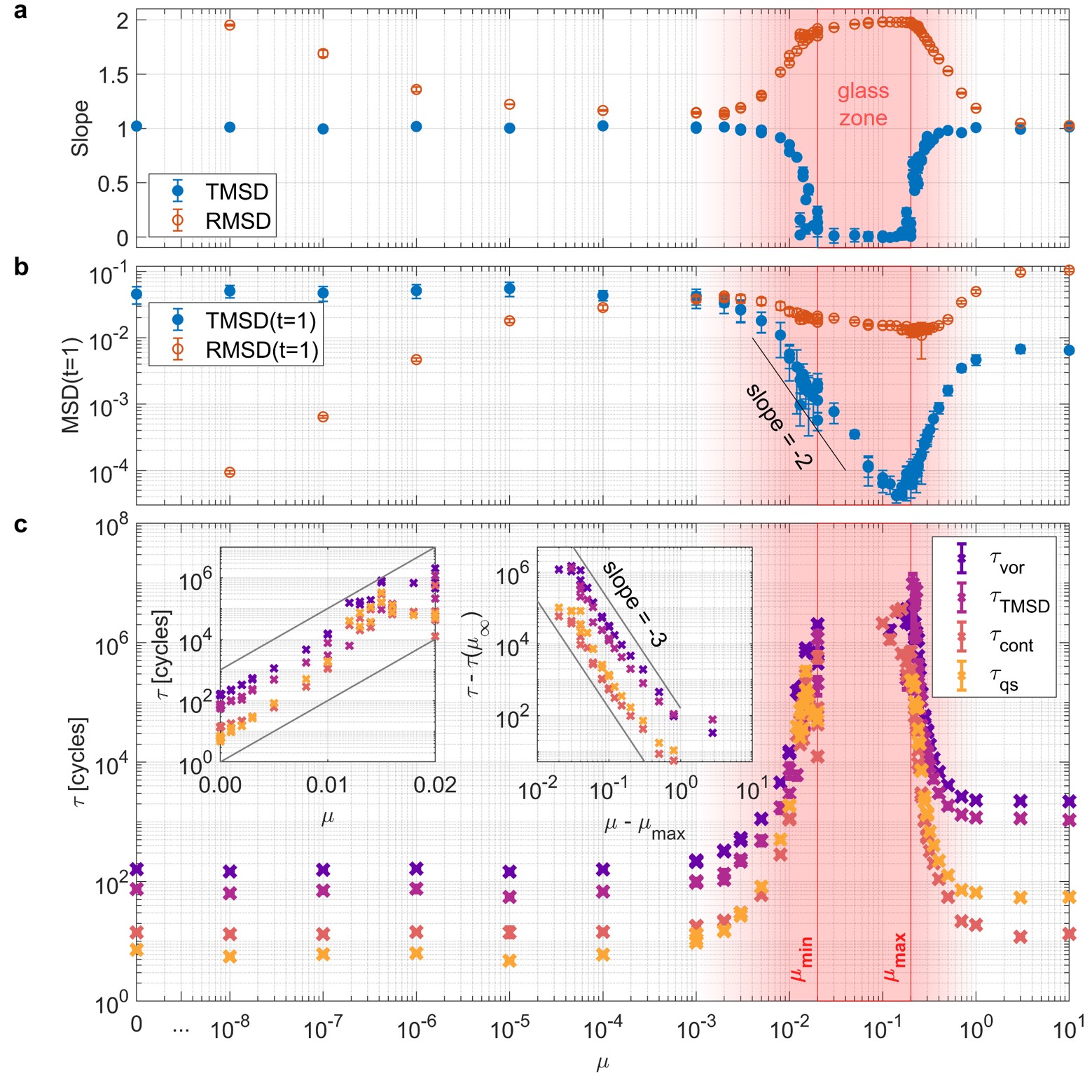}
\caption{
{\bf Relaxation times for the translational and rotational degrees of freedom as a function of friction coefficient:}
(a) Slope of the TMSD and RMSD at long times, obtained from the last three decades of the curves in Fig.~\ref{fig_ED_TMSD_RMSD}. Error-bars indicate 95\% confidence interval.
(b) Value of TMSD and RMSD at $t=1$. The straight line indicates a power-law with exponent -2. Error-bars indicate the standard deviation.
(c) $\mu$-dependence of various relaxation times: $\tau_\text{cont}$, the time for a particle to disconnect from a neighbor with which it was in contact at $t=0$; $\tau_\text{vor}$, the characteristic time for a particle to have a change in its (radical) Voronoi neighborhood; $\tau_\text{qs}$, the decay-time of the self-overlap function (threshold 0.3$d_s$), and $\tau_\text{TMSD}$, the time for the TMSD to reach $0.1$. The $t-$dependence of the corresponding time correlation functions is shown in Extended Data Fig.~\ref{fig_ED_time_dependence}. Error-bars indicate 95\% confidence interval.
Left Inset: Same data for small $\mu$ in a log-lin plot. The gray straight lines are exponentials with a slope 200. 
Right Inset: Same data for large $\mu$ as a function of $\mu-\mu_{\rm max}$. The straight lines have slope -3.0. The vertical red lines in the three panels indicate $\mu_{\rm min}$ and $\mu_{\rm max}$.}
\label{fig_1}
\end{figure}

Since the time dependence of the MSDs does not show for all $\mu$ a diffusive behavior, the diffusion constant cannot be used to quantify the dynamics. To this aim we therefore monitor MSD($t=1$) as well as the long-time slope of the MSD in a double-logarithmic plot. Figure~\ref{fig_1}(a) shows that for $\mu \lesssim 2 \cdot 10^{-3}$ the TMSD has a slope of 1.0 and then drops quickly to a value that is basically zero, i.e., the dynamics changes from diffusive to caged. This caging regime extends from around $\mu_{\rm min}=2\cdot 10^{-2}$ to $\mu_{\rm max}=0.2$ and will be referred to as ``glass-zone'' since the TDOF are completely frozen. For $\mu> \mu_{\rm max}$ the slope shoots back up to $1.0$ and remains constant. The exponent for the RMSD decreases from 2.0 at $\mu=0$ (ballistic motion) to a value slightly above $1$ at $\mu \approx 10^{-3}$. If $\mu$ is increased further, the slope rises quickly and becomes 2.0 in the glass-zone, demonstrating that some of the particles undergo free rotation~\cite{zhao2022PRX, zhao2022FP}.
For $\mu \geq \mu_{\rm max}$, the slope decreases again and becomes 1.0, signaling diffusive rotation. 

The short time dynamics is quantified by the value of the MSD at $t=1$, Fig.~\ref{fig_1}(b). At small friction the TMSD($t=1$) is large and stays constant for $\mu \leq 10^{-3}$, i.e., in this range of friction the short time translational dynamics is independent of $\mu$ and the system is similar to 2D glass-formers~\cite{flenner2015NC}. Upon approaching $\mu_{\rm min}$, TMSD$(t=1)$ starts to drop and reaches a minimum at around $\mu=0.15$. This decrease, which is directly related to a shrinking cage-size, is compatible with a power-law with an exponent of -2.0.
Beyond $\mu_{\rm max}$, TMSD$(t=1)$ increases by two orders of magnitude, signaling the acceleration of the dynamics, before reaching a plateau. 
The  RMSD$(t=1)$ shows qualitatively the same trend, although the $\mu$-dependence is much weaker. Note that the first signs that friction influences significantly the dynamics are seen at around $\mu=10^{-3}$. This surprisingly small value implies that the dynamics of real granular systems cannot be described theoretically if friction is not taken into account.

Fig.~\ref{fig_1}(c) presents the $\mu$-dependence of different characteristic times $\tau$ (defined in the caption). All times increase exponentially quickly upon approaching $\mu_{\rm min}$  (left inset of panel (c)), beyond which $\tau$ cannot be determined reliably anymore. In contrast to this we find for $\mu \geq \mu_{\rm max}$ that $\tau(\mu)$ is given by a power-law with a critical point at $\mu_{\rm max}$, right inset of panel (c), hinting the relation to the isostaticity point of the pack~\cite{shundyak2007PRE}. The presence of these two different growth laws indicates that the mechanisms leading to the increase of $\tau$ are distinct.\\

\newpage

\begin{figure}[ht]

\makebox[\textwidth][c]{\includegraphics[width=18cm]{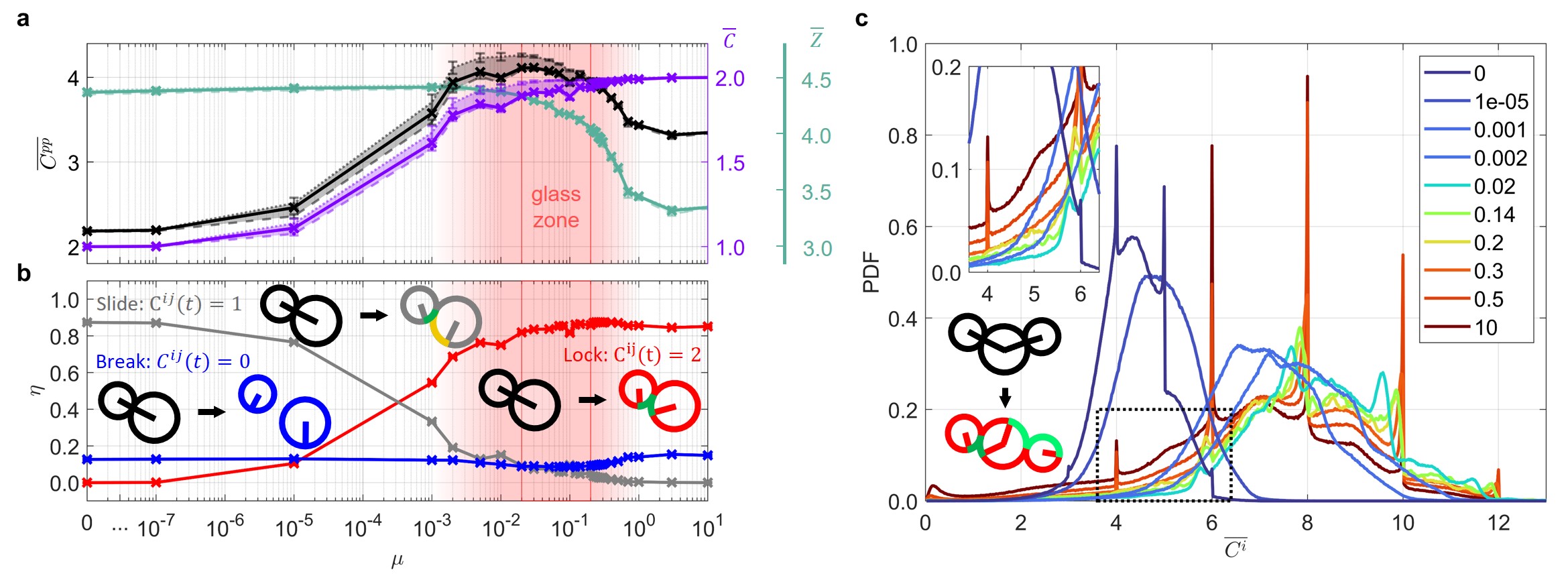}}
\caption{{\bf Distributions of averaged number of constraints as a function of friction coefficient:} (a) Effective number of constraints per contact, $\overline{C}$, number of constraints per particle, $\overline{C^{pp}}$, and the averaged number of contacts, $\overline{Z}$. Solid lines show the time-averaged results, while dashed lines and dotted lines show results at $\gamma=0$ and $\gamma=0.05$, respectively. Error-bars indicate the standard deviation. (b) Cartoons show three different states of a contact: Sliding, breaking, locked. The curves show the probability that during a cycle a contact is in a given state. (c) Distribution of the time-averaged number of constraints per particle, $\overline{C^i}$. Inset: Zoom on the dotted rectangular region. Cartoon: Chain-like motion corresponding to the peak at $\overline{C^i}=4$.}
\label{fig_2}
\end{figure}

{\bf Dynamic constraints:}
For many-body systems the slowing down of the dynamics is often related to a change in structural quantities like the packing fraction or coordination number~\cite{binder2011,gotze2008}. Fig.~\ref{fig_2}(a) shows that the cycle averaged coordination number, $\overline{Z}$, is slightly above 4.0 if $\mu$ is small, while for large friction it slightly exceeds 3.0, thus compatible with the values expected for a 2-dimensional static pack of particles with infinite elastic moduli~\cite{wyart2005ADP}. 
That $\overline{Z}$ (as well as the packing fraction, see Extended Data Fig.~\ref{fig_ED_phi_rattler_ratio_tau}) decreases simply monotonically with $\mu$ indicates that static quantities like $\overline{Z}$ are not able to rationalize the re-entrance dynamics and one needs also to consider the dynamics on the contact level. Hence we probe the probabilities $\eta$ that during a cycle a contact i) breaks, ii) persists and slides, or iii) is completed frictional locked, see cartoons in Fig.~\ref{fig_2}(b). That panel demonstrates that $\eta_\text{\rm break}$ is low and basically independent of $\mu$, except for a small rise at the largest $\mu$.  $\eta_\text{slide}$ decreases monotonically to zero with increasing friction, while $\eta_\text{lock}$ increases monotonically and eventually saturates, which demonstrates that with increasing $\mu$ the contact dynamic crosses over from sliding to rolling.

Maxwell's idea which relates the number of constraints to the static mechanical rigidity of a structure can be generalized to the dynamic case by defining for each contact in the pack a cycle-averaged number of constraints as follows. 
During a cycle, one monitors the contact between particles $i$ and $j$ and defines $C^{ij}(t)$, the associated number of constraints for this contact by distinguishing the three cases introduced above:
i) The contact breaks, which corresponds to a normal force $|{\bf F}_n^{ij}|=0$ and hence the contact is not constrained, i.e., $C^{ij}(t)=0$; 
ii) The contact has $|{\bf F}_n^{ij}|>0$ with a tangential force $|{\bf F}_t^{ij}|> \mu |{\bf F}_n^{ij}|$, thus is sliding, giving a constraint of 1; iii) If $|{\bf F}_t^{ij}| < \mu |{\bf F}_n^{ij}|$ one has slipless rotation which corresponds to two constraints. The cycle-averaged dynamic constraint of the contact, $\overline{C^{ij}}$, is thus given by

\begin{equation}
\overline{C^{ij}}=\frac{1}{T}\int_0^T C^{ij}(t) dt, \qquad  {\rm where} \quad C^{ij}(t) = \left\{
\begin{aligned}
&0 \quad {\rm if } \quad |{\bf F}_n^{ij}(t)|=0\\
&1 \quad {\rm if } \quad |{\bf F}_n^{ij}(t)|>0 \text{ and } |{\bf F}_t^{ij}(t)| = \mu |{\bf F}_n^{ij}(t)|\\
&2 \quad {\rm if } \quad |{\bf F}_n^{ij}(t)|>0 \text{ and } |{\bf F}_t^{ij}(t)| < \mu |{\bf F}_n^{ij}(t)|.\\
\end{aligned}
\right.
\label{eq:1}
\end{equation}
($T$ is the cycle time and details on the calculation are in the Methods.) 
The number of constraints acting on particle $i$ is given by $\overline {C^i}=\sum_{j} \overline{C^{ij}}$ and this quantity will indicate whether or not the motion of the particle is blocked.  The system average of $\overline {C^i}$, $\overline{C^{pp}}=N^{-1} \sum_{i=1}^N \overline {C^i}$, gives the average number of constraints per particle and shows, Fig.~\ref{fig_2}(a), that for small friction the number of constraints is just above 2.0, rises gently towards 4.0 if $\mu$ approaches $\mu_{\rm min}$, before decreasing again at the largest $\mu$.
The $\mu$-dependence of $\overline{C^{pp}}$ can be rationalized by making the mean-field approximation $\overline{C^{pp}} \approx \overline{Z} \cdot \overline{C} /2$, where $\overline{C}$ is the system- and time-averaged number of constraints per contact, shown in Fig.~\ref{fig_2}(a). ($\overline{C^{pp}}$ is compared with mean-field in the SI.) One observes that at small friction $\overline{C^{pp}}$ grows because $\overline{C}$ increases, while $\overline{Z}$ stays constant. The product reaches a maximum at around $\mu_{\rm min}$ and stays constant up to around $\mu_{\rm max}$ since the slight decrease of $\overline{Z}$ is compensated by the increase of $\overline{C}$. For $\mu > \mu_{\rm max}$, $\overline{Z}$ drops quickly, i.e., the structure becomes more open, while $\overline{C}$ remains constant, resulting in a quick drop of $\overline{C^{pp}}$, ensuing an acceleration of the dynamics.
This result is thus strong evidence that the non-monotonic behavior of the relaxation dynamics can be understood directly from the $\mu$-dependence of the number of dynamic constraints on the particles.

The PDF of $\overline{C^i}$, Fig.~\ref{fig_2}(c), allows to get insight on its average, $\overline{C^{pp}}$. For $\mu=0$ there are two narrow peaks at 4 and 5, corresponding to particles that have, respectively, four and five frictionless contacts, i.e., $\overline{C^{ij}}=1$. 
For somewhat larger $\mu$, they are absorbed into the broader peak which shifts to the right, i.e., most of the contacts are at least temporarily locked. Upon approaching the glass-region, $\mu=0.02$, the PDF starts to show again several narrow peaks, $\overline{C^{ij}}\approx 7.8$ and 9.6, 
and an increase of $\mu$ to 0.3 makes these two peaks sharp and move to 8 and 10, respectively, and three others appear, at 4, 6, and 12, corresponding to particles that have 2, 3,...,6 neighbors with locked contacts, i.e., $\overline{C^{ij}}=2$. Note that the peak at 4 is the last to emerge, at around $\mu_{\rm max}$ (Inset). This demonstrates that for $\mu>\mu_{\rm max}$ a typical local structure of the system is a linear chain in which each particle has exactly 2 permanently locked contacts, see cartoon in Fig.~\ref{fig_2}(c), indicating that spatial correlations of constrained contacts play an important role for the relaxation dynamics and below we will see that this is indeed the case. \\

{\bf Spatial structure of constraints:}
We now probe the nature of the cooperative motion, often observed in other glassy materials~\cite{binder2011,berthier2011}, in our system and relate it to the local structure and the constraints. For this we compare the spatial map of $\overline{C^{ij}}$ with the one of $D_{\rm min}^2$, the non-affine displacement of a particle with respect to its nearest neighbor shell~\cite{falk1998PRE}.

\begin{figure}[ht]
\vspace*{-1.5cm}
\includegraphics[width=12cm]{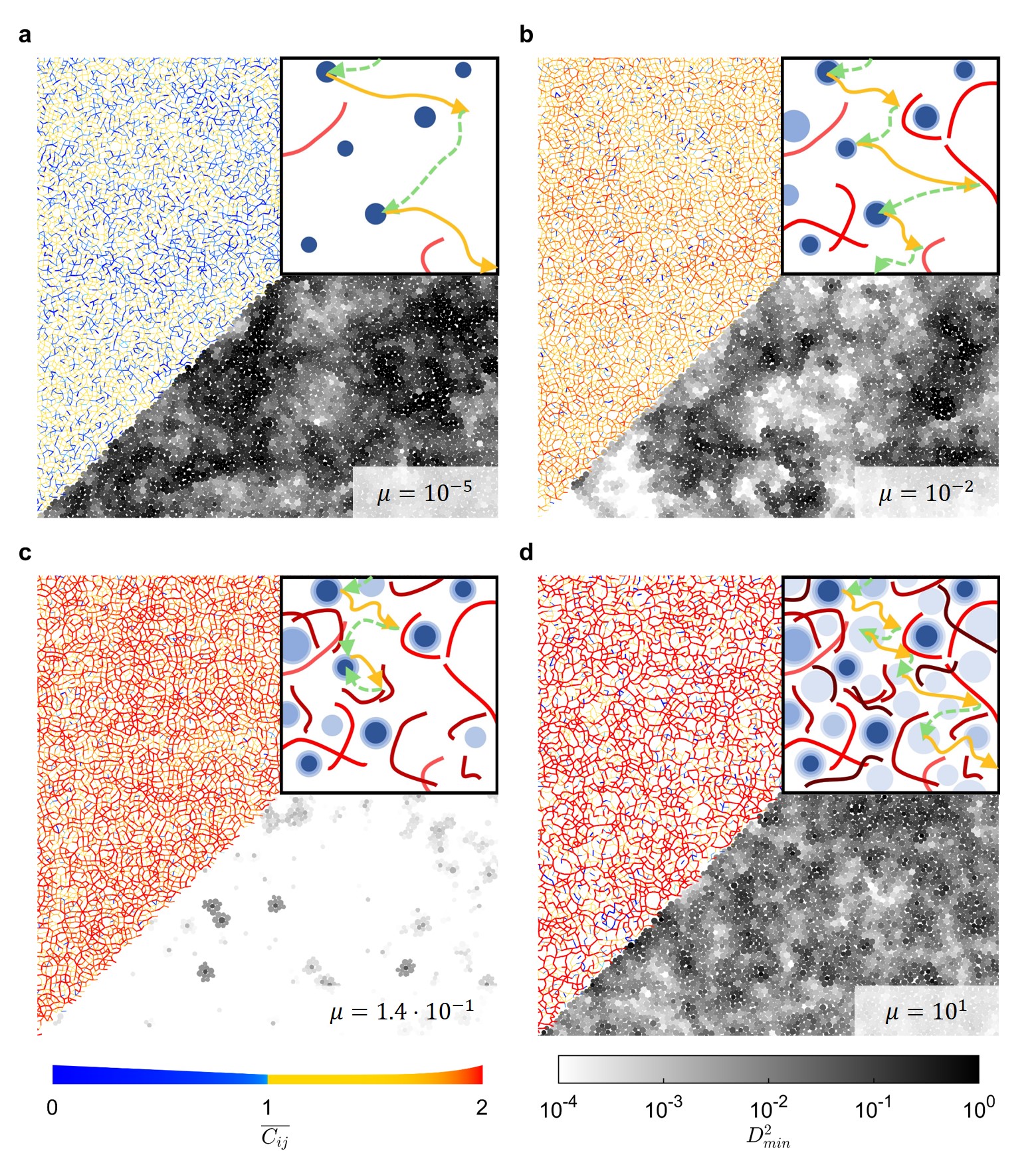}
\caption{
{\bf Spatial structure of non-affine displacement and of dynamic constraints on contacts for different $\mu$.} Upper left part of the panels show the network of cycle-averaged contacts, $\overline{C^{ij}}$, with the color code given at the bottom left. Lower right part of the panels show the non-affine displacement after one cycle, grey scale code at bottom right. See main text for a discussion of these panels. Insets show schematically the motion of the system in configuration space. Dark blue: Mechanically stable regions where the particles have typically a coordination number of 4. Light blue: regions with typically coordination number 3. Orange/green arrows correspond to the forward/backward motion of the cycle. Red lines are barriers that are incompatible with the dynamic constraints and hence the system avoids them. Insets: (a) Many stable regions have coordination number 4 and the absence of constraints allows the system to move easily in configuration space. (b) The constraints make that the motion becomes hindered resulting to a slowed down relaxation dynamics. (c) The number of constraints is so high that motion is reversible and relaxation suppressed. Formation of first zones with coordination number 3. (d) High friction allows the system to access mechanically stable state in which particles have a typical coordination number of 3. These states allow the system to explore new pathways and hence to relax.}
\label{fig_3}
\end{figure}

\begin{figure}[h]
\makebox[\textwidth][c]{\includegraphics[width=18cm]{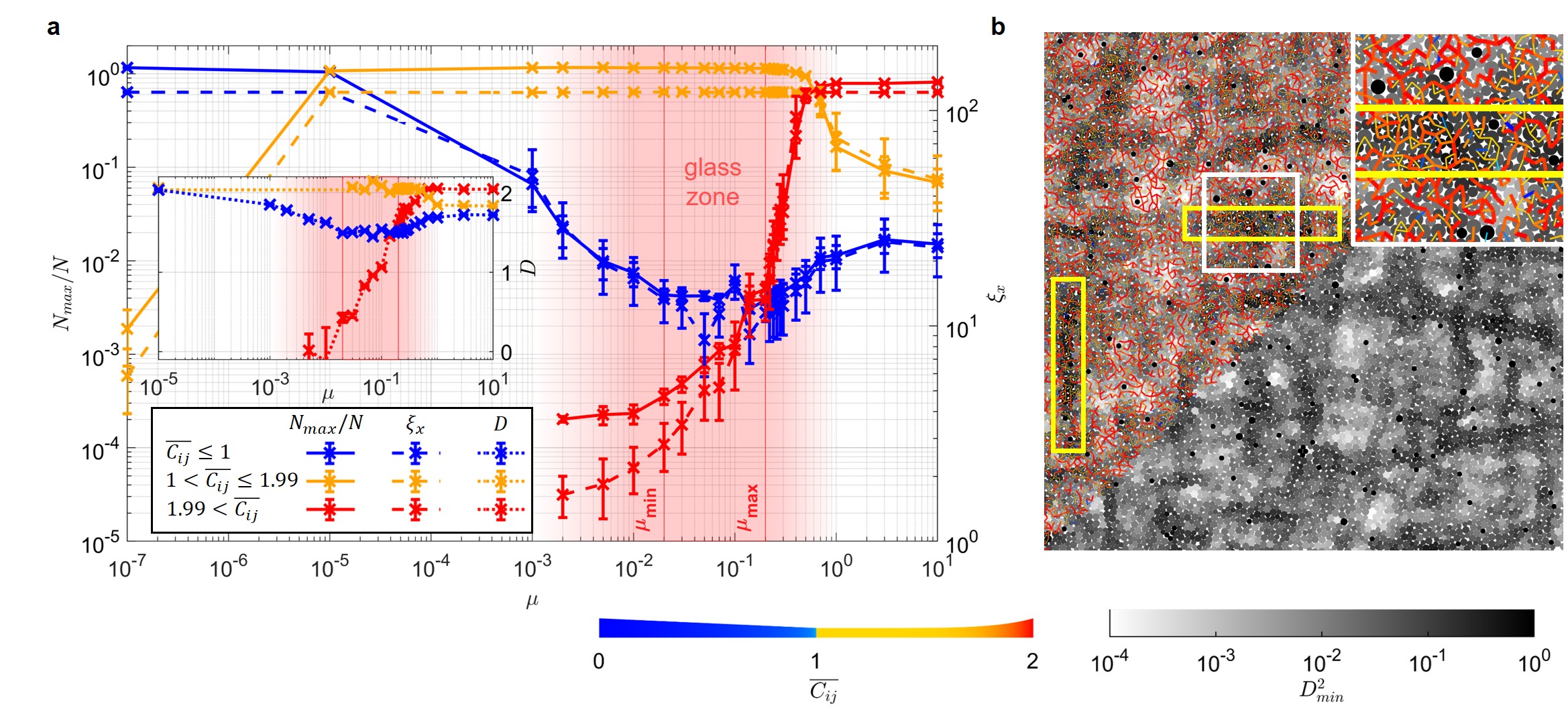}}
\caption{{\bf Percolation probability and length scales of the contact network.}
(a): The number of contact in the largest connected cluster of $\overline{C^{ij}}$ (normalized by the total number of contacts), full lines, the extension of this cluster in the $x$-direction, dashed lines, and (Inset) the fractal dimension of this cluster, dotted lines. The three colors correspond to the three ranges of contacts given in the legend. Error-bars indicate the standard deviation. 
(b): Network of contacts and non-affine displacement field at maximal strain; $\mu=1.4 \cdot 10^{-1}$. The micro-shear bands where $D^2_{\rm min}$ is high after a quarter cycle (yellow boxes) are basically reversible once the shear cycle is completed, see Fig.~\ref{fig_3}(c). Inset: Zoom on the white square area.
}
\label{fig_4}
\end{figure}

The upper left parts of Fig.~\ref{fig_3} show the maps of $\overline{C^{ij}}$ for increasing $\mu$, and Fig.~\ref{fig_4}(a) presents the cluster analysis. We find that this structure transforms from a percolating $\overline{C^{ij}}\leq 1$ network at $\mu=10^{-5}$, Fig.~\ref{fig_3}(a), to a percolating $1<\overline{C^{ij}}\leq 1.99$ one when $\mu$ approaches $\mu_{\rm min}$, Fig.~\ref{fig_3}(b), while contacts $\overline{C^{ij}}> 1.99$ are scarce. This demonstrates that the dynamic slowing down is not related to completely locked contacts but instead to the percolating network with intermediate $\overline{C^{ij}}$. Even deep in the glass zone, Fig.~\ref{fig_3}(c), the fully locked contacts do not yet percolate, but they do so at the highest $\mu$, Fig.~\ref{fig_3}(d).
This percolating network consists of chain-like structures pictured in Fig.~\ref{fig_2}(c), and therefore can deform more easily in a collective manner.

The network of $\overline{C^{ij}}$ determines how the system can relax, described by $D_{\rm min}^2$ in the lower right part of the panels. At small $\mu$, Fig.~\ref{fig_3}(a), $D_{\rm min}^2$ is spatially very heterogeneous, since the percolating network $\overline{C^{ij}}\leq 1$ constraints the dynamics only weakly, thus giving rise to the heterogeneous dynamics found in other glass-forming systems. For $\mu=10^{-2}$, Fig.~\ref{fig_3}(b), the dynamical heterogeneities increase strongly, quantified via $\chi_4$ in Extended Data Fig.~\ref{fig_ED_chi4}, and the average magnitude of $D_{\rm min}^2$ decreases, since the dynamics is now confined by the percolating $1<\overline{C^{ij}}\leq 1.99$ network. Inside the glass zone, Fig.~\ref{fig_3}(c), the constraints are so strong that most of the particles cannot change anymore their neighbors and hence the zones with significant $D_{\rm min}^2$ become small islands. If $\mu$ exceeds $\mu_{\rm max}$, Fig.~\ref{fig_3}(d), zones with large $D_{\rm min}^2$ reappear, but in contrast to the case of small $\mu$, they form a multitude of small clusters, which are embedded in the highly constrained contact network. \\

{\bf Dynamic constraints in configuration space:} The influence of the constraints on the dynamics can be understood via the propagation of the system in configuration space, illustrated by the cartoons on the upper right corner of the panels in Fig.~\ref{fig_3}. At very low $\mu$, the system can freely explore the landscape between blue stable areas since there are only very few constraints. With increasing $\mu<\mu_{\rm min}$, the exploration is hindered by barriers from increasing frictional constraints, red lines, therefore the motion becomes more collective and slower.

Inside the glass region, the system is likely to occupy places surrounded by many more barriers, therefore the relaxation ceases. To accommodate the strain, the system forms narrow micro-shear bands, mainly oriented horizontally and vertically, in which the contacts slide because they have a small $\overline{C^{ij}}$, Fig.~\ref{fig_4}(b). The particles inside these bands have a large $D_{\rm min}^2$ at maximal strain, but most motion is reversible. Thus, although trapped near a local minimum of the free energy landscape, the system has self-organized its structure permitting it to have a high elasticity.

When $\mu>\mu_{max}$, the number of stable structures increases, since friction lowers the minimal coordination number, allowing new relaxation channels to circumvent the barriers. This is accomplished by the deformation of the open highly-constrained network, which allows the particles enclosed to undergo a non-affine motion to relax. Some of these cells in the network become unstable during the cycle, i.e., the local bubble bursts, leading ultimately to the relaxation of the entire network.

The evolution of the structure in configuration space demonstrates the dual role of friction: Increasing $\mu$ results in a higher $\overline{C}$ which means more dynamic barriers, thus slower relaxation. At around $\mu_{\rm max}$ friction allows the creation of many stable states with low $\overline{Z}$, facilitating relaxation via new pathways of deformation and reconstruction of the network.\\

{\bf Outlook:} Our results demonstrate that the number of dynamic constraints is a good indicator of the relaxation dynamics, and provide more insight than the often studied force chain network, see Extended Data Figs.~\ref{fig_ED_pdf_forces}, \ref{fig_ED_force_network}. This implies that in frictional systems the dynamics is governed by the distance of a contact from its Coulomb boundary and not by the magnitude of its force, a conclusion that holds for all materials, irrespective of their friction coefficients (hydrogels, grains, gears particles,..), thus opening the door to tune the relaxation dynamics by controlling the friction coefficient via methods like surface chemical treatment~\cite{Kim2023SM}. Our approach allows us to treat the dynamics of granular material and frictionless hard-sphere glass formers within the same framework, which will permit to unify the physics of various disordered systems. The approach of dynamic constraints to study dynamics is very general and hence it can be adapted to other soft matter systems like suspensions, deformable cells, and biological tissues~\cite{vandernaald2024NP,hannezo2022TCB,atia2018NP}. As many systems (in and out of equilibrium) have complex free energy landscapes, and often have a dynamics subject to time-varying constraints, applications in domains like the traveling salesman problem, financial markets, and ecology network~\cite{bigras2008DO,charbonneau2023}, are expected to be feasible.

\clearpage
\newpage

{\bf Methods:}

{\it Model and Simulations:} The contact force between particles is Hertzian, and the tangential component
is subject to a Coulomb cutoff~\cite{ohern2002PRL,bares2017PRE}. Thus the total force between two particles $i$ and $j$ is given by 
$$
{\bf F}(r)=\sqrt{R_i R_j/(R_i+R_j)} \sqrt{\delta}\left[(k_n\delta -m_{\rm eff} \gamma_n v_n){\bf n} -(k_t \Delta s_t + m_{\rm eff} \gamma_t v_t) {\bf t}\right].
$$ 
Here ${\bf n}$ and ${\bf t}$ are, respectively, the normal and tangent vectors at the contact, and the overlap between the particles is given by $\delta=R_i+R_j-r$, where $R_i$ is the radius of the particle. $v_n$ and $v_t$ are the relative speed of the particles, projected on ${\bf n}$ and ${\bf t}$, respectively, and $m_{\rm eff}$ is the effective mass. $\Delta s_t$ is the displacement between two particles in the tangential direction. The absolute value of $k_t \Delta s_t$ is subject to cutoff at $\mu k_n\delta$. We parameters we use are $k_n=2\cdot10^8 \text{g}\cdot \text{cm}^{-1} \cdot \text{s}^{-2}$, $\gamma_n=11,900 \text{cm}^{-1} \cdot \text{s}^{-1}$, $k_t=\frac{2}{7}k_n$, and $\gamma_t=\frac{1}{2}\gamma_n$. Simulations were carried out using the LAMMPS software~\cite{LAMMPS} with a time step size of $4\cdot 10^{-5}$s. The simulation shear box, shown in Inset of Fig.~\ref{fig_ED_TMSD_RMSD}(b), consists of four amorphous walls to avoid heterogeneous crystallization of the sample from the boundary and also to weaken the global convection effect that can occur along smooth boundaries. Similar to an experimental setup for sheared systems~\cite{kou2017NATURE}, the bottom wall (pink) moves sinusoidally in the horizontal direction with a shear amplitude of $\gamma_0=0.05$ and a strain rate of $\dot{\gamma} = \text{0.1 s}^{-1}$, while the left and the right walls (blue) move accordingly. The top  wall (green) applies a pressure of $P=98100 \text{g}\cdot\text{s}^{-2}$ on the pack and it is constrained to move only vertically. 
This gives an inertia number $I=\dot{\gamma}\bar{d}/\sqrt{P/\rho}$ is $3.6\cdot 10^{-4}$, where $\bar{d}$ is the mean diameter of the particles, i.e., we are in the quasi-static shear regime~\cite{midi2004EPJESM}. Typically we used for each values of $\mu$ 1-4 independent samples and did up to $9\cdot 10^5$ cycles to reach the steady state. See Extended Data Table~I for details.

{\it Defining locked contacts:} In practice we define a contact to be non-slipping, i.e., ``locked'', if the tangential force is inside the Coulomb-cone within a relative accuracy $\varepsilon$, i.e., $|{\bf F}_t|< \mu (1-\varepsilon)|{\bf F}_n|$, where we have chosen $\varepsilon=10^{-6}$. To improve the statistics for obtaining the results for the probabilities shown in Fig.~\ref{fig_2}(b) we averaged over 700 cycles. To calculate the integral in Eq.~\ref{eq:1} we divided the cycle into 100 equal intervals.

\clearpage
\newpage

\printbibliography[notkeyword={methods}]

\clearpage
\newpage

\clearpage
\newpage

{\bf Extended Data:}\\

\renewcommand{\figurename}{Extended Data Fig.}
\setcounter{figure}{0}

\vspace*{20mm}

\begin{table}[h]
\begin{tabular}{||l|l||l|l||l|l||l|l||}
\hline
$\mu$       & cycles($10^4$) & $\mu$    & cycles($10^4$)  & $\mu$   & cycles($10^4$) & $\mu$   & cycles($10^4$) \\ \hline
0        & 6            & 0.01  & 14, 48        & 0.1  & 6, 6         & 0.26 & 48, 72       \\ \hline
$10^{-8}$ & 6            & 0.012 & 30            & 0.12 & 6            & 0.28 & 12, 48       \\ \hline
$10^{-7}$ & 6            & 0.013 & 36, 84        & 0.14 & 6            & 0.3  & 12, 72, 84   \\ \hline
$10^{-6}$ & 6            & 0.014 & 24, 60        & 0.16 & 24,24        & 0.32 & 6            \\ \hline
$10^{-5}$ & 6            & 0.015 & 66, 72        & 0.18 & 18, 24       & 0.35 & 6            \\ \hline
0.0001   & 6            & 0.016 & 24, 30        & 0.2  & 23, 24, 24   & 0.4  & 6            \\ \hline
0.001    & 6, 6         & 0.018 & 18            & 0.21 & 24, 60       & 0.5  & 6            \\ \hline
0.002    & 18, 30       & 0.02  & 2, 12, 18, 54 & 0.22 & 24, 72       & 0.7  & 6            \\ \hline
0.003    & 6, 12        & 0.03  & 2             & 0.23 & 48, 90       & 1    & 6            \\ \hline
0.005    & 16, 24       & 0.05  & 2             & 0.24 & 30, 60, 66   & 3    & 6            \\ \hline
0.008    & 6            & 0.07  & 2, 6          & 0.25 & 54           & 10   & 6            \\ \hline
\end{tabular}

\caption{{\bf Table of simulation cycle numbers for each $\mu$:} Each number represent a run with independent random initial condition.}
\end{table}

\begin{figure}[t!]
\makebox[\textwidth][c]{\includegraphics[width=18cm]{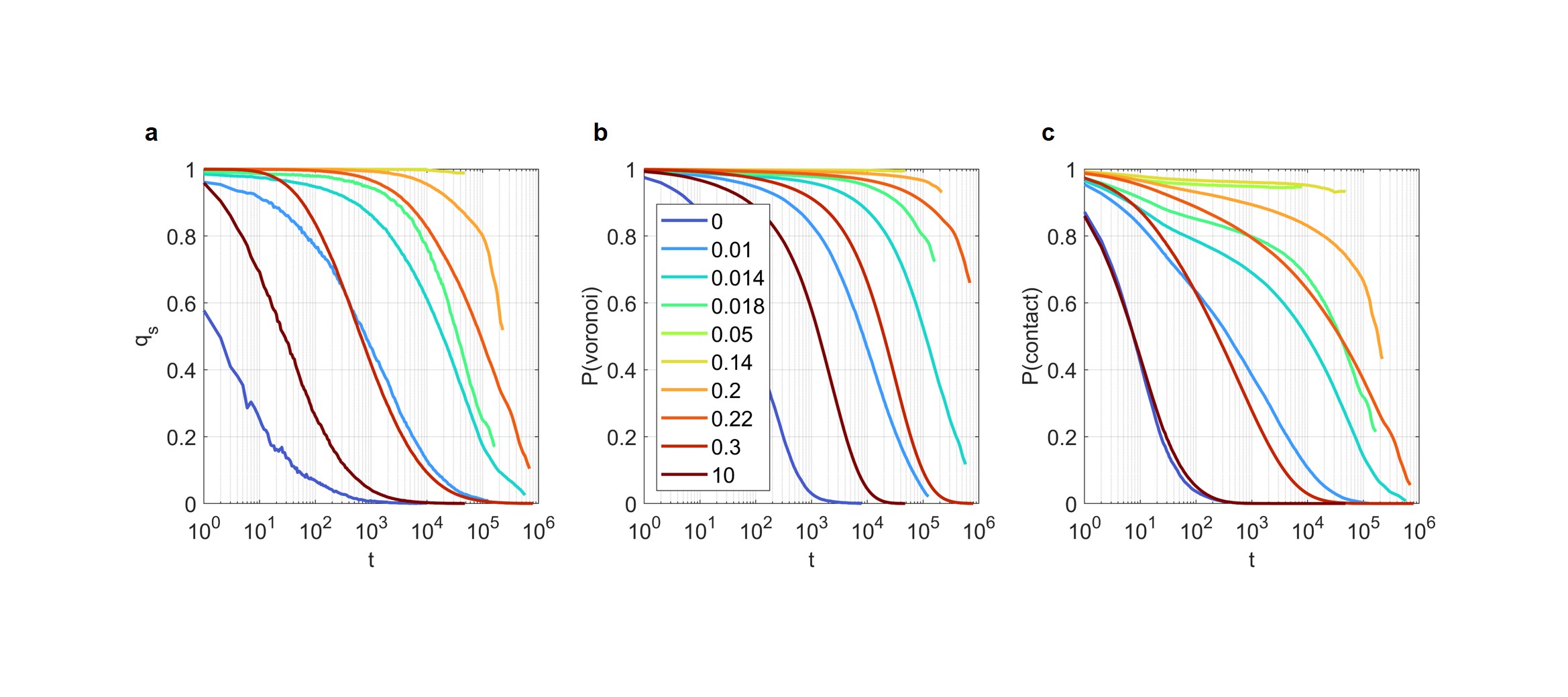}}
\caption{{\bf Time dependence of persistence probabilities:} (a) Time dependence of the self-overlap (defined using a threshold of 0.3$d_s$). (b) Probability that a particle that at $t=0$ was in a given Voronoi cell, is still in the same cell at time $t$. (c) The probability that at time $t$ a contact that was present at time $t=0$ is still intact.
}
\label{fig_ED_time_dependence}
\end{figure}

\begin{figure}[t!]
\includegraphics[width=\textwidth]{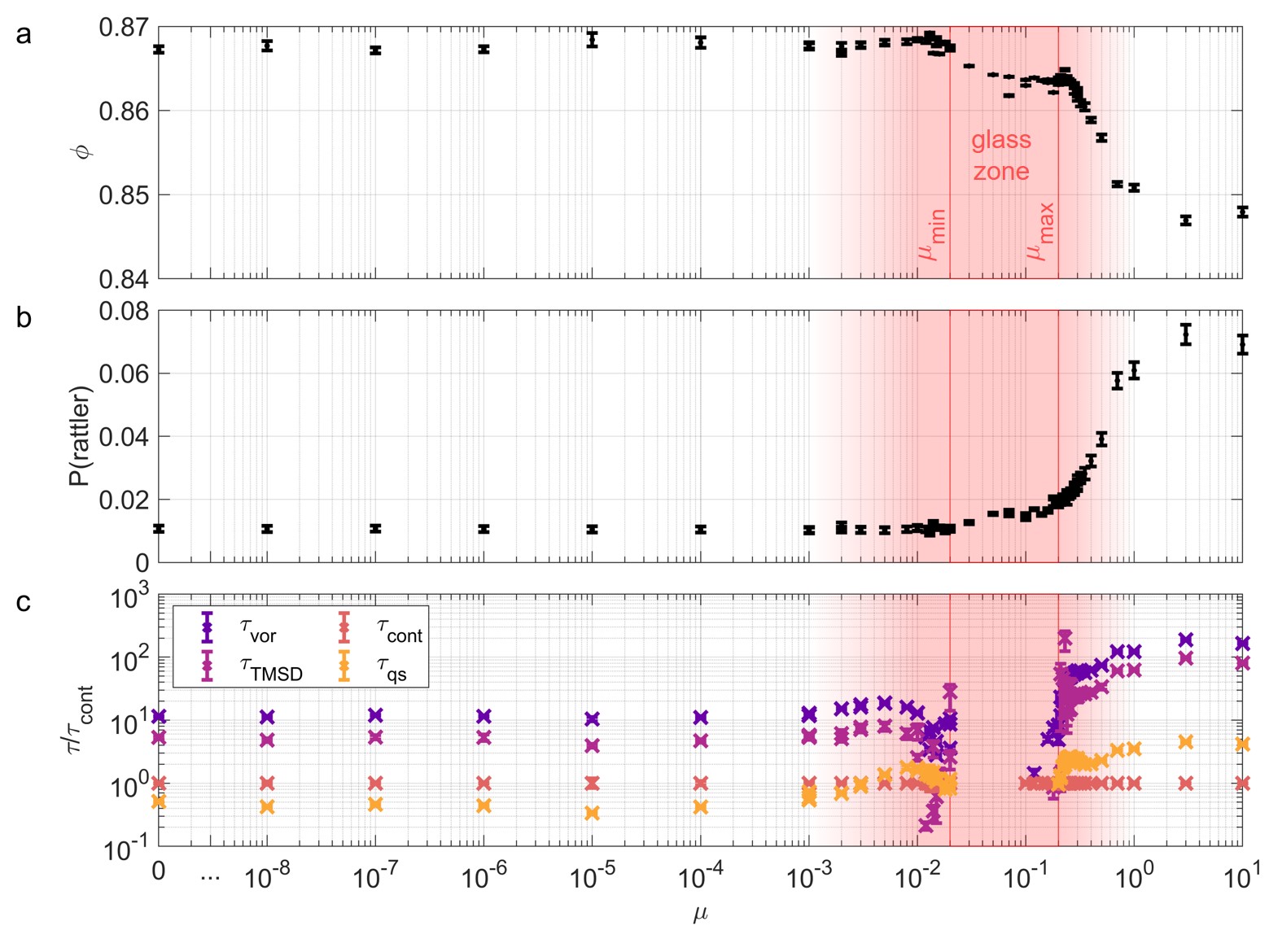}
\caption{{\bf Packing fraction, concentration of rattlers, relaxation times:} (a) Packing fraction as a function of $\mu$. $\phi$ decreases continuously with increasing $\mu$, making it unlikely that the re-entrant dynamics is only an effect of the packing fraction. At the upper bound of the glass-regime, i.e., $\mu_{\rm max}$, $\phi$ drops quickly because the increased friction allows to stabilize states in which the particles have a low coordination number. (b) Probability that a particle is a rattler, i.e, particles with fewer than 2 contacts. This probability is low ($\approx 1$\%) at small and intermediate fraction, i.e., as long as the typical coordination number is larger than 4. At high $\mu$ this probability grows significantly due to the formation of the open network structure discussed in the main text. Error-bars in (a) and (b) indicate the standard deviation. (c) $\mu$-dependence of the relaxation times, presented in Fig.~1 of the main text, normalized by $\tau_{\rm cont}$, the time it takes for a particle to break a contact. Error-bars indicate 95\% confidence interval.}
\label{fig_ED_phi_rattler_ratio_tau}
\end{figure}

\begin{figure}[t!]
\includegraphics[width=12cm]{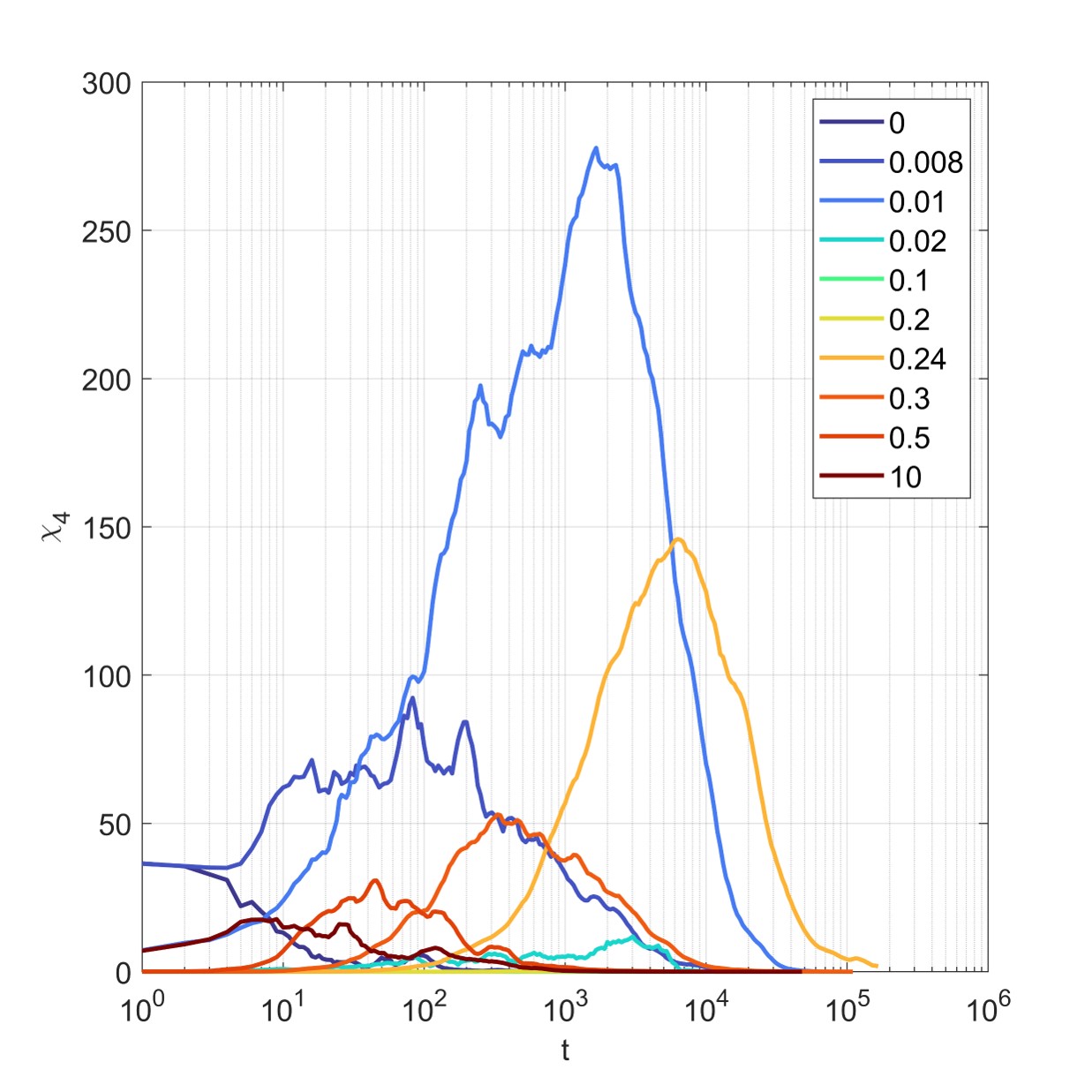}
\caption{{\bf Time dependence of $\chi_4$ for different values of $\mu$:} The dynamic susceptibility (self part) is used to quantify the strength of the dynamical heterogeneities in the system and it has been calculated from the variance of the overlap function. $\chi_4$ is relatively small for small friction and increases rapidly if $\mu$ approaches the glass-regime. Inside the glass-regime $\chi_4$ is basically zero, since the system hardly relaxes. For $\mu \gtrsim \mu_{\rm max}$ it is again large and decreases steadily with increasing friction. At the same time also the time scale at which $\chi_4(t)$ peaks is non-monotonic in $\mu$. These results demonstrate that the system has pronounced dynamical heterogeneities that peak at the boundary of the glass-zone.
}
\label{fig_ED_chi4}
\end{figure}

\begin{figure}[t!]
\includegraphics[width=\textwidth]{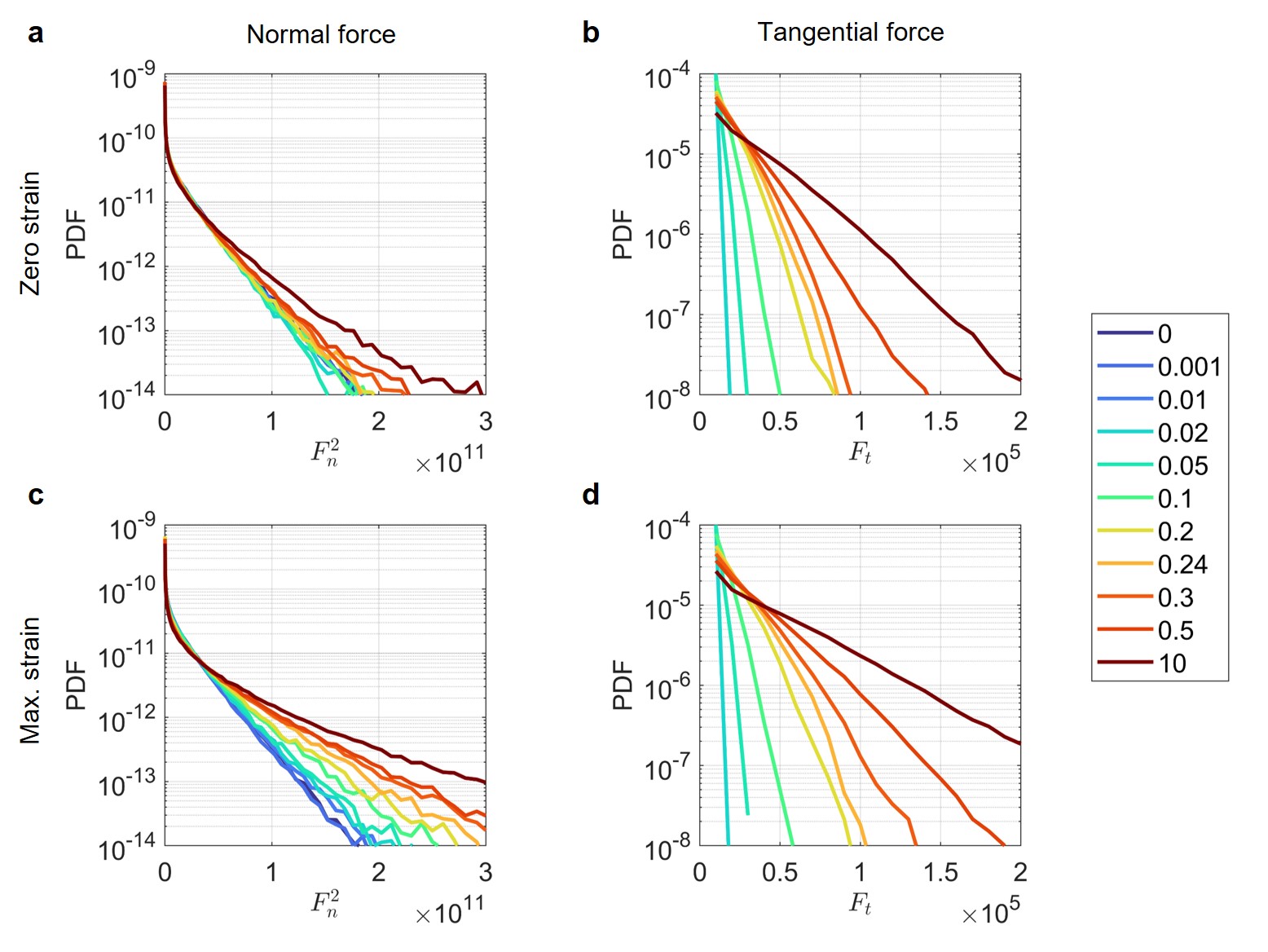}
\caption{{\bf Probability distribution function of the forces:} Panels (a) and (b) show the PDF of forces at strain $\gamma=0$ and panels (c) and (d) shows the ones at $\gamma=0.05$. The PDF's for ${\bf F}_n$ show a Gaussian decay for all values of $\mu$. The width of the Gaussian increases in a monotonic manner as a function of $\mu$ at $\gamma=0.05$, and fall on a master curve for $\gamma=0$. The PDFs of ${\bf F}_t$ follow an exponential distribution with an increasing width. These results indicate that the forces are not able to rationalize the re-entrant dynamics of the system.
} 
\label{fig_ED_pdf_forces}
\end{figure}
  
\begin{figure}[t!]
\makebox[\textwidth][c]{\includegraphics[width=18cm]{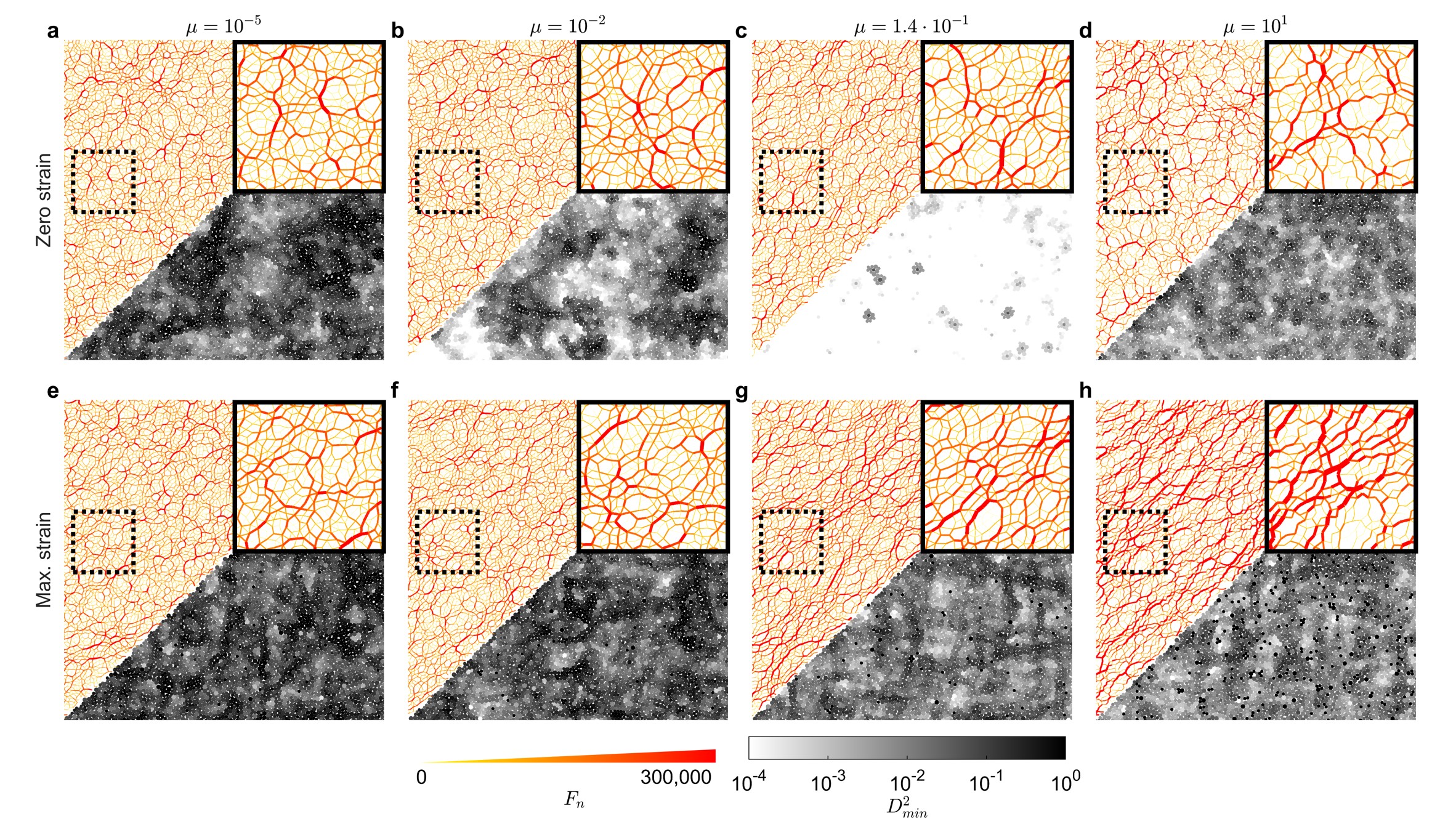}}
\caption{{\bf Network of the normal forces at the contacts:} The upper left part of the panels shows the magnitude of the normal component of the force at the contacts while the lower right part shows the non-affine displacement of the particles. Scales are given at the bottom of the figure. Panels (a)-(d) are at strain $\gamma=0$ for different $\mu$ and panels (e)-(f) at strain $\gamma=0.05$. No evident non-monotonic evolution is seen in the force chains when $\mu$ is increased.
}
\label{fig_ED_force_network}
\end{figure}

\clearpage
\newpage

\printbibliography[keyword={methods}]

\clearpage
\newpage

{\bf Supplementary Information}\\[-4mm]

{\bf Mean field approximation}
Here we present the comparison between the mean field result and the one of the simulation.

The number of constraints related to a contact between particle $i$ and $j$ at time $t$ can be written as
\begin{equation*}
    C^{ij}(t)=1\cdot\vmathbb{1}^{ij}(t)+2\cdot\vmathbb{2}^{ij}(t) \quad,
\end{equation*}

\noindent
where $\vmathbb{1}^{ij}(t)$ and $\vmathbb{2}^{ij}(t)$ are functions describing the state of this contact at time $t$: $\vmathbb{1}^{ij}(t)=1$ and $\vmathbb{2}^{ij}(t)=0$ if the contact is sliding; $\vmathbb{1}^{ij}(t)=0$ and $\vmathbb{2}^{ij}(t)=1$ if the contact is locked; $\vmathbb{1}^{ij}(t)=0$ and $\vmathbb{2}^{ij}(t)=0$ if the contact is broken.

The time-averaged constraint of this contact is 
\begin{equation*}
    \overline{C^{ij}}=\frac{1}{T}\int^T_0 C^{ij}(t)dt \quad .
\end{equation*}

We define the time-averaged constraint per particle as

\begin{equation*}
\overline{C^{pp}}=\frac{1}{2N}\sum_{i, j}\overline{C^{ij}}=\frac{1}{2NT}\sum_{i, j}\int^T_0 (\vmathbb{1}^{ij}(t)+2\cdot\vmathbb{2}^{ij}(t))dt \quad ,
\end{equation*}

\noindent
where $N$ is the particle number.

The system averaged contact number per particle at time $t$, i.e., the average coordination number of a particle, is given by

\begin{equation*}
    Z(t) = \frac{1}{N}\sum_{i, j} \vmathbb{1}^{ij}(t)+\vmathbb{2}^{ij}(t)\quad ,
\end{equation*}

\noindent
and thus the time-averaged number of contacts is

\begin{equation*}
    \overline{Z} = \frac{1}{NT}\sum_{i, j} \int^T_0 (\vmathbb{1}^{ij}(t)+\vmathbb{2}^{ij}(t))dt \quad .
\end{equation*}

Now we define the effective number of constraints per contact at time $t$ as

\begin{equation*}
    C(t)=\frac{\sum_{i, j}\vmathbb{1}^{ij}(t)+2\cdot\vmathbb{2}^{ij}(t)}{\sum_{i, j}\vmathbb{1}^{ij}(t)+\vmathbb{2}^{ij}(t)} \quad,
\end{equation*}

\noindent
where the numerator is two times the total number of contacts at time $t$. The $\mu-$dependence of the ensemble averaged value of $C(t)$ at $t=0$ and $t=T/4$, which corresponds to strain $\gamma=0$ and 0.05, respectively,  are shown in Fig. 2a as a dotted and a dashed line, respectively.

The time-averaged effective number of constraints per contact is thus

\begin{equation*}
    \overline{C} = \frac{1}{T}\int^T_0\frac{\sum_{i, j}\vmathbb{1}^{ij}(t)+2\cdot\vmathbb{2}^{ij}(t)}{\sum_{i, j}\vmathbb{1}^{ij}(t)+\vmathbb{2}^{ij}(t)}dt \quad .
\end{equation*}

We now consider different way of averaging, by first calculating the averaged number of total number of constraints in a cycle, and the averaged number of total contact time in a cycle, and subsequently their ratio: 

\begin{equation*}
    \overline{C'} = \frac{\frac{1}{T}\sum_{i, j}\int^T_0(\vmathbb{1}^{ij}(t)+2\cdot\vmathbb{2}^{ij}(t))dt}{\frac{1}{T}\sum_{i, j}\int^T_0(\vmathbb{1}^{ij}(t)+\vmathbb{2}^{ij}(t))dt} \quad .
\end{equation*}

Note that

\begin{equation*}
    \frac{1}{2}\overline{C'}\cdot\overline{Z}  = \overline{C^{pp}} \quad .
\end{equation*}

Now we compare the difference between $\overline{C}$ and $\overline{C'}$. 
The $\mu-$dependence of the ensemble average of $Z(t)$ at $=0$ and $t=T/4$, corresponding, respectively, to $\gamma=0$ and  $0.05$, are shown in Fig. 2a, dotted and dashed line, respectively. Note that these two curves are basically identical, i.e., the $t$-dependence of $Z(t)$ is weak. This suggests that one can make the approximation
\begin{equation*}
\overline{Z}\approx Z(t) \quad .
\end{equation*}

We have 
$$\overline{C} = \frac{1}{T}\int^T_0\frac{\sum_{i, j}\vmathbb{1}^{ij}(t)+2\cdot\vmathbb{2}^{ij}(t)}{NZ(t)}dt\approx\frac{1}{NT\overline{Z}}\int^T_0\sum_{i, j}(\vmathbb{1}^{ij}(t)+2\cdot\vmathbb{2}^{ij}(t))dt$$

and the last expression is just $\overline{C'}$, i.e., one has
\begin{equation*}
    \overline{C}\approx\overline{C'}
\end{equation*}
Thus we conclude that
\begin{equation*}
    \overline{C^{pp}} \approx \frac{1}{2}\overline{C}\cdot\overline{Z}.
\end{equation*}

\newpage

\clearpage
\newpage

\end{document}